# Frequency Reference Stability and Coherence Loss in Radio Astronomy Interferometers Application to the SKA

Bassem Alachkar[1,2], Althea Wilkinson[1] and Keith Grainge[1]

[1]*Jodrell Bank Centre for Astrophysics*
*School of Physics & Astronomy*
*University of Manchester*
*Manchester, UK*
[2]*bassem.alachkar@manchester.ac.uk*

The requirements on the stability of the frequency reference in the Square Kilometre Array (SKA), as a radio astronomy interferometer, are given in terms of maximum accepted degree of coherence loss caused by the instability of the frequency reference. In this paper we analyse the relationship between the characterisation of the instability of the frequency reference in the radio astronomy array and the coherence loss. The calculation of the coherence loss from the instability characterisation given by the Allan deviation is reviewed. Some practical aspects and limitations are analysed.

## 1. Introduction

The basic principle of radio astronomy interferometry relies on combining the received signals of the array sensors in a coherent manner. To achieve good coherence of the array, the frequency references at the array sensors must be synchronised to a common reference. In a connected radio interferometry array, this can be achieved by a synchronised frequency dissemination system which delivers frequency reference signals to the receptors, (e.g. Schediwy *et al.*, 2017 and Wang *et al.*, 2015). The local frequency reference signals are used for sampling and/or for down-converting the frequency of the radio astronomy signals. Any instability of the frequency references causes loss of coherence in the array. Therefore, the requirements on the stability of the synchronisation system can be defined in such a way as to limit the coherence loss caused by frequency instability. The stability of a frequency reference is usually characterised by the Allan deviation, while the coherence between two signals is calculated by the correlation integral of the signals. This paper analyses the characterisation and evaluation of frequency distribution system instability and the relationship with the coherence loss in a radio astronomy array. The application of this analysis is based on the coherence requirements of the Square Kilometre Array (SKA) (Dewdney et al., 2009).

In the SKA, the Synchronisation and Timing (SAT) system provides frequency and timing signals to the receptors and other parts of the telescope (Grainge *et al., 2017*). These signals are required to have certain levels of stability and accuracy, specified by the SKA requirements. The basic SAT stability and accuracy requirements are derived from the astronomical science requirements, covering the radio astronomy functions of the telescope for imaging, pulsar searching and timing, and Very Long Baseline Interferometry (VLBI). The relevant SKA requirements for frequency distribution are given in terms of coherence loss, so that the maximum coherence loss accepted for frequencies up to 13.8 GHz and for intervals of 1 s or 60 s is limited to 2% in total. These requirements are driven by the need for coherence over the correlator integration time (approximately 1 second) and the time for in-beam calibration (1 minute). They are expressed in terms of the coherence loss caused by the phase difference between the two frequency signals delivered at two receptors, or in other words, per baseline.

In section 2, we give the definition of coherence and coherence loss in radio astronomy interferometry, and link it to the phase noise produced by the frequency reference. In section 3, the model of phase noise is presented and the instability of a frequency reference is analysed. Allan deviation characterisation is considered as it is the common tool used for frequency reference characterisation. The main practical cases and aspects of the phase noise are presented. In section 4, the relationship between the frequency instability and resulting coherence loss is analysed. In section 5, some practical aspects are analysed, and examples of simulation results are given. Section 6 gives the conclusions of this analysis.

## 2. Coherence Loss in a Radio Astronomy Interferometer

The coherence loss in radio astronomy interferometry caused by frequency reference instability, especially in VLBI, has been analysed by several authors such as (Rogers & Moran 1981), (Thompson, Moran & Swenson, 2004) and (Kawaguchi, 1983).

The coherence function for an integration time $T$ is defined as in (Rogers & Moran 1981) and (Thompson, Moran & Swenson, 2004):

$$C(T) = \left| \frac{1}{T} \int_0^T exp[i\varphi(t)]dt \right|, \tag{1}$$

where $\varphi(t)$ is the phase difference between the two stations of the interferometer. There is no loss of coherence if the value of $C(T)$ is 1. The coherence loss $L_C$ is

$$L_C = 1 - \sqrt{\langle C^2(T) \rangle} \tag{2}$$

where $\sqrt{\langle C^2(T) \rangle}$ is the root mean-square of the coherence $C(T)$ for an integration time $T$.

The mean-square value of $C(T)$ is

$$\langle C^2(T) \rangle = \langle \frac{1}{T^2} \int_0^T \int_0^T exp\{i[\varphi(t) - \varphi(t')]\}dt \, dt' \rangle \tag{3}$$

If $\varphi$ is a Gaussian random variable, we find:

$$\langle C^2(T) \rangle = \frac{1}{T^2} \int_0^T \int_0^T exp\left[\frac{-\sigma^2(t,t')}{2}\right] dt \, dt' \tag{4}$$

where $\sigma^2(t, t')$ is the variance of the phase difference between the two instants $t$ and $t'$.

$$\sigma^2(t, t') = \langle [\varphi(t') - \varphi(t)]^2 \rangle \tag{5}$$

If we assume that $\sigma^2$ is stationary and depends only on $\tau = t' - t$ then it could be written as:

$$\sigma^2(\tau) = \langle [\varphi(t + \tau) - \varphi(t)]^2 \rangle \tag{6}$$

The mean squared coherence can then be written as:

$$\langle C^2(T) \rangle = \frac{2}{T} \int_0^T \left(1 - \frac{\tau}{T}\right) exp\left[\frac{-\sigma^2(\tau)}{2}\right] d\tau \tag{7}$$

## 3. Phase Noise and Instability of the Frequency Reference

### 3.1. *The Frequency Signal Model*

The phase ϕ(t) of an oscillator or frequency reference is defined with reference to another frequency reference which is usually considered to be of accuracy and stability acceptable for the application. The instantaneous frequency is the derivative of the phase ϕ(t) = 2π υ₀t + φ(t):

$$v(t) = v_0 + \delta v(t) \tag{8}$$

$v_0$ is the nominal frequency of the frequency reference, and the frequency deviation is:

$$\delta v(t) = \frac{1}{2\pi} \frac{d\varphi(t)}{dt} \tag{9}$$

The instantaneous fractional frequency deviation is:

$$y(t) = \frac{\delta v(t)}{v_0} \tag{10}$$

The deviations of the frequency are described as deterministic systematic deviations or random deviations. In general, the model of these deviations is given by the following:

$$y(t) = y_0 + Dt + \epsilon(t) \tag{11}$$

where the systematic deviations are covered by the frequency offset $y_0$ and the linear frequency drift $Dt$, while the random deviations are covered by $\epsilon(t)$. The systematic deviations are usually small and have small effects on the phase deviation in short periods of time. In the model given by equation (11), systematic deviations of higher order are neglected.

## 3.2. *Allan Variance*

Allan variance is the most common statistical function used to characterise and classify frequency fluctuations of a frequency reference (Allan, 1966) (Rutman, 1978). It overcomes convergence and non-stationarity difficulties that might appear for some phase noise cases (Lindsey & Chie, 1976).
The fractional frequency fluctuation averaged over the time interval $\tau$ is:

$$\bar{y}_k = \frac{1}{\tau} \int_{t_k}^{t_k+\tau} y(t) dt = \frac{\varphi(t_k+\tau) - \varphi(t_k)}{2\pi v_0 \tau} \tag{12}$$

Using the sample variance of $\bar{y}_k$, as a measure of the instability of oscillators or frequency references is not always convenient because it does not converge in many cases of phase noise. This is due to the low frequency behaviour. The two-sample or Allan variance $\sigma_y^2(\tau)$ is generally accepted as a measure of instability of the oscillators or frequency references as it overcomes the convergence problem for most practical cases. The Allan variance is defined as follows:

$$\sigma_y^2(\tau) = \frac{\langle (\bar{y}_{k+1} - \bar{y}_k)^2 \rangle}{2} \tag{13}$$

The Allan deviation is $\sigma_y(\tau)$. This is usually used to characterise the random deviations $\epsilon(t)$ which are related to the noise in the frequency. The Allan deviation is not affected by the frequency offset $y_0$ because of the differential operation in its definition Eq. (13). The drift $Dt$ results in an additive contribution of $\sigma_y^2(\tau) = \frac{1}{2} D^2 \tau^2$ to the Allan variance.

**Phase noise types**

In the frequency domain the phase and frequency fluctuations are characterised by their power spectral densities. If $S_\varphi(f)$ and $S_{\delta v}(f)$ are the one-sided (the Fourier frequency goes from 0 to $\infty$) spectral densities of the phase and frequency fluctuations respectively, they are related by the following relationship:

$$S_{\delta v}(f) = f^2 S_\varphi(f) \tag{14}$$

The power spectral density $S_y(f)$ of the fractional frequency deviation is related to $S_{\delta v}(f)$ and $S_\varphi(f)$ by:

$$S_y(f) = \frac{1}{v_0^2} S_{\delta v}(f) = \frac{f^2}{v_0^2} S_\varphi(f) \tag{15}$$

For most practical applications the spectral densities of the random frequency fluctuations are considered to follow a power law model (Barnes et al., 1971) and (Rutman, 1978):

$$S_y(f) = \sum_{\alpha=-2}^{+2} h_\alpha f^\alpha \tag{16}$$

for $0 \leq f \leq f_h$ where $f_h$ is an upper cut-off frequency. A cut-off frequency is related to a limitation in the bandwidth in the reference frequency generator itself, as in the case of the phase locked loops (PLLs), or in the measurement systems.

For $\alpha = -2$ to $+2$, the types of noise are random-walk frequency, flicker frequency, white frequency, flicker phase, and white phase noise, respectively.

Two simple cases which are of practical importance are the cases of white phase noise and white-frequency noise. Figures 5 and 8 in section 5 present examples of these types.

**White-phase noise**

The power spectrum density of the fractional frequency deviation is given by the form $S_y(f) = h_2 f^2$.

In this case and for $\tau \gg \frac{1}{2\pi f_h}$, the Allan variance is given by, see (Rutman, 1978):

$$\sigma_y^2(\tau) = \frac{3 h_2 f_h}{(2\pi)^2 \tau^2} \tag{17}$$

**White-frequency noise**

The power spectrum density of the fractional frequency deviation is given by the form $S_y(f) = h_0$.

In this case and for $\tau \gg \frac{1}{2\pi f_h}$, the Allan variance is given by, see (Rutman, 1978):

$$\sigma_y^2(\tau) = \frac{h_0}{2\tau} \tag{18}$$

## 4. Coherence Loss and Allan deviation

The purpose of this section is to review the relationship between the coherence loss caused by the random component of the phase instability of the frequency reference in an interferometer and the characterisation of this reference by the Allan deviation.

Using (13) and (12), $\sigma_y^2(\tau)$ can be written as follows:

$$\sigma_y^2(\tau) = \frac{1}{2} \frac{\langle [\varphi(t+2\tau) - 2\varphi(t+\tau) + \varphi(t)]^2 \rangle}{(2\pi \nu_0 \tau)^2} \tag{19}$$

Using (6) and (19) and following (Rogers & Moran 1981), (Thompson, Moran & Swenson, 2004), the variance of the phase difference $\sigma^2(\tau)$ is expressed as:

$$\sigma^2(\tau) = \frac{1}{2} (2\pi \nu_0 \tau)^2 [\sigma_y^2(\tau) + \sigma_y^2(2\tau) + \sigma_y^2(4\tau) + \sigma_y^2(8\tau) + \cdots ] \tag{20}$$

provided the series converges.

Then (7) is written as:

$$\langle C^2(T) \rangle = \frac{2}{T} \int_0^T \left(1 - \frac{\tau}{T}\right) \exp\{-(\pi \nu_0 \tau)^2 [\sigma_y^2(\tau) + \sigma_y^2(2\tau) + \sigma_y^2(4\tau) + \cdots ]\} d\tau \tag{21}$$

The last equation can be simplified in the cases of white phase noise and white frequency noise.

**White-phase noise**

In this case, the power spectrum density of the fractional frequency deviation is given by the form $S_y(f) = h_2 f^2$.

In this case and for $\tau \gg \frac{1}{2\pi f_h}$, where $f_h$ is the upper cutoff frequency, the Allan variance is given by (17).

And the mean squared coherence can be evaluated from (21):

$$\langle C^2(T)\rangle = \exp(-h_2 f_h \upsilon_0^2) \tag{22}$$

Hence the coherence loss in this case is:

$$L_C = 1 - \sqrt{\exp(-h_2 f_h \upsilon_0^2)} \tag{23}$$

The factor $h_2 f_h$ in the equation (23) can be extracted from the Allan deviation function given by the equation (17).

In fact, in the case of 'white' phase noise, the power spectrum density of the phase fluctuation is a constant $h_2 \upsilon_0^2$, and over the band of frequency $[0, f_h]$ the power of this 'white' signal is $f_h h_2 \upsilon_0^2$ which is $R_\varphi(0)$ the variance of the white signal $\varphi(t)$. In reality, it is not an ideal white signal as the power spectrum density is considered constant over only a band limited by the cut-off frequency $f_h$. An ideal white signal does not exist in nature. From equation (6), the variance of the phase fluctuation is $\sigma^2(\tau) = \langle[\varphi(t+\tau) - \varphi(t)]^2\rangle = 2[R_\varphi(0) - R_\varphi(\tau)]$, where $R_\varphi(\tau)$ is the autocorrelation and $R_\varphi(0)$ is the variance and the power of the phase deviation $\varphi(t)$. As $\varphi(t)$ is a "white" signal within the frequency band $[0, f_h]$ and for $\tau \gg \frac{1}{2\pi f_h}$ we know that $R_\varphi(\tau) \approx 0$, then $\sigma^2(\tau) = 2R_\varphi(0)$. The mean square coherence given in (7) becomes in this case:

$$\langle C^2(T)\rangle = exp[-R_\varphi(0)] \tag{24}$$

which is the same equation as (22), as the power of the signal is $R_\varphi(0) = f_h h_2 \upsilon_0^2$.

The coherence, in this case, depends only on the mean square of the phase deviation φ, and does not depend on the integration time T. Figure 1 shows the coherence loss as a function of the root mean square of the phase deviation $\sqrt{R_\varphi(0)}$ in the case of 'white' phase noise.

**White-frequency noise**

The power spectrum density of the fractional frequency deviation is given by the form $S_y(f) = h_0$.

In this case and for $\tau \gg \frac{1}{2\pi f_h}$, the Allan variance is given by (18).

In this case the mean squared coherence is given by the following, (Thompson, Moran & Swenson, 2004):

$$\langle C^2(T)\rangle = \frac{2(e^{-aT}+aT-1)}{a^2T^2} \tag{25}$$

where

$$a = \pi^2 \upsilon_0^2 h_0 \tag{26}$$

Hence the coherence loss in this case is:

$$L_C(T) = 1 - \sqrt{\frac{2(e^{-aT}+aT-1)}{a^2T^2}} \tag{27}$$

The factor $h_0$ and therefore $a$ in equations (26) and (27) can be extracted from the Allan deviation function given in equation (18). The coherence loss in this case depends on the integration time $T$.

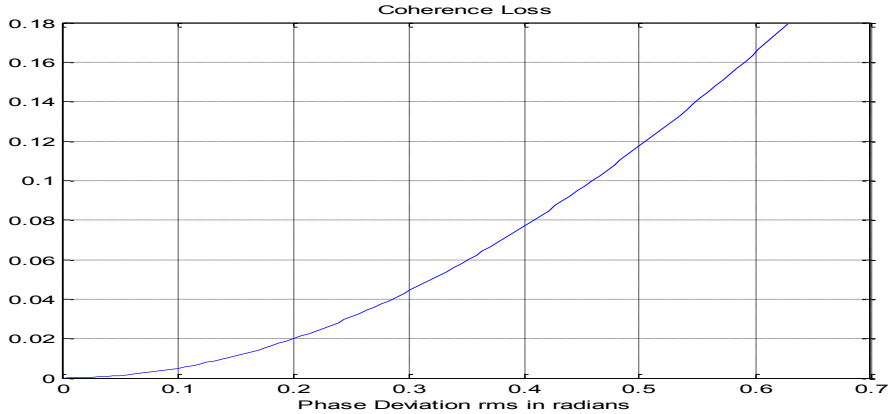

Fig. 1. The coherence loss as a function of the root mean square of the phase deviation in the case of 'white' phase noise.

## 5. Verification of the Coherence Loss Requirements

**Allan deviation for different phase noise types**

Given the Allan deviation of the frequency reference, we can calculate the coherence loss between two receptors (i.e. for a baseline) of an interferometer caused by the instability of the frequency reference, by using equation (21). Equation (21) can be applied in the general case. If the Allan deviation characterisation shows white phase noise or white frequency noise behaviour, the coherence loss can be calculated directly using equations (23) and (27). In the general case, when the Allan deviation curve suggests that the noise falls in neither specific phase noise regime, least squares methods can be applied to the data to calculate a 'best' fit of the function $\sigma_y^2(\tau)$, which can then be used in equation (21).

**The phase difference between two receptors**

The loss of coherence due to the instability of the frequency references between two stations or receptors of the interferometer array depends on the fluctuations of the phase difference $\varphi(t)$ of the two frequency references. The stability characterisation of one frequency reference is usually given with reference to another much more stable frequency reference. In the case of the frequency dissemination systems, the phase difference considered in the stability characterisation (e. g. Allan deviation) is between the frequency reference delivered at the receptor and a common central frequency reference. Different link lengths cause different levels of instability. The Allan variance or the variance in phase difference between the frequency references in the two receptors is the sum of the corresponding variances between each of these points and the central reference, assuming that the variations on the two links causing the instability are independent.

**Frequency conversion of local oscillator or sampling clock reference**

The frequency reference at the astronomy receptor might be used as a local oscillator to be mixed with the astronomical signals for converting the frequency before sampling, or used directly as a reference for the sampling clock of the analog-to-digital convertor.

In the first case, the phase deviation of the local oscillator is added directly to the phase of the astronomical signal. In equation (21), the frequency $\upsilon_0$ is the local oscillator frequency, which is the frequency reference and the Allan deviation is the Allan deviation of this frequency reference.

In the second case, when the frequency reference is used to provide the sampling clock of sampling frequency $F_S$, we will see that in equation (21) the frequency $\upsilon_0$ is the astronomical observation frequency and the Allan deviation is again the Allan deviation of the frequency reference.

Consider a frequency component $x(t) = A\, exp[i(2\pi f_0 t)]$ of the astronomical observation signal. The sampled signal $x_S(n)$, resulting from sampling $x(t)$ at the sampling instants $t_n = \frac{n}{F_S}$ generated by an ideal frequency reference, is: $x_S(n) = A\, exp\left[i(2\pi \frac{f_0}{F_S} n)\right]$. The phase deviation $\Delta\varphi$ in the real frequency reference will cause a shift in the sampling instant so that it becomes: $t'_n = \frac{n}{F_S} + \Delta T$, where $\Delta T = \frac{\Delta\varphi}{2\pi F_S}$. The sampled signal becomes: $x'_S(n) = A\, exp\left[i\left[\left(2\pi \frac{f_0}{F_S} n\right) + \frac{f_0}{F_S}\Delta\varphi\right]\right]$.

This is equivalent to introducing a phase of $\frac{f_0}{F_S}\Delta\varphi$ to the signal $x(t)$. If $\sigma_y(\tau)$ is the Allan deviation corresponding to the phase deviation $\Delta\varphi$ of a frequency reference with a nominal frequency $F_S$, then the same value $\sigma_y(\tau)$ will be the Allan deviation corresponding to the phase deviation $\frac{f_0}{F_S}\Delta\varphi$ of a frequency reference with a nominal frequency $f_0$.

We conclude that the phase deviation introduced by sampling a signal of frequency $f_0$ with a sampling frequency reference with Allan deviation $\sigma_y(\tau)$ will result in a loss of coherence equivalent to the coherence loss caused by an oscillator of frequency $f_0$ and Allan deviation $\sigma_y(\tau)$, and therefore equation (21) is applicable with $v_0 = f_0$, the observation frequency, and $\sigma_y(\tau)$ is the Allan deviation of the frequency reference.

**Coherence loss calculation from phase measurements:**
Numerically, the coherence loss is calculated from the discrete phase sequence $\varphi_d(n) = \varphi(nT_S)$, which results from the sampling of the continuous phase signal $\varphi(t)$. $T_S$ is the sampling period. The integral (1) is calculated by the sum:

$$\hat{C}(T) = \left|\frac{1}{N}\sum_{n=1}^{N} exp[i\varphi_d(n)]\right| \tag{28}$$

$\hat{C}(T)$ is the estimation of the coherence on an interval of length $T$, where $T = NT_S$.
The sequence of the measured phase signal is divided into $K$ segments of length $T$, the coherence $\hat{C}_k(T)$ is calculated on each segment $k$, and then the estimation of the coherence loss $\hat{L}_C$ is calculated as following:

$$\hat{L}_C(T) = 1 - \sqrt{\frac{1}{K}\sum_{k=1}^{K}\hat{C}_k(T)^2} \tag{29}$$

**Coherence Loss calculation from Allan deviation characterisation:**

From a set of values of $\sigma_y^2$ given for a number of averaging times $\tau_m$, the Allan deviation function $\sigma_y^2(\tau)$ is estimated by a least squares fitting. $\widehat{\sigma_y^2}(\tau)$ is the least squared fitting function, which is defined by a limited number of parameters. Then the coherence loss is estimated numerically by the equation:

$$\langle\widehat{C^2(T)}\rangle = \frac{2}{N}\sum_{n=0}^{N-1}\left(1 - \frac{n}{N}\right) exp\{-(\pi v_0 n\Delta\tau)^2[\widehat{\sigma_y^2}(n\Delta\tau) + \widehat{\sigma_y^2}(2n\Delta\tau) + \widehat{\sigma_y^2}(4n\Delta\tau) + \cdots]\} \tag{30}$$

where $T = N\Delta\tau$.

**Cases of white frequency noise and white phase noise**
The coherence requirements for the SKA limit the coherence loss to 2% at an integration time of 60 seconds. If the Allan deviation characterisation of the frequency distribution system shows that the system has either white frequency noise or white phase noise, then with the integration time, the value of the Allan deviation at $\tau = 1$ second is sufficient to determine the coherence loss.
In figures 2 and 3, the coherence loss at integration time of 1 s and 60 s is given in the cases of white frequency noise and white phase noise as a function of the Allan deviation value at $\tau = 1$ second between two receptors, for an operating frequency 13.8 GHz. The coherence loss is independent of the integration time in case of the white phase noise and it increases with the integration time in the case of white frequency noise. Figure 4 gives the coherence loss as a function of the integration time for different values of Allan deviation at $\tau = 1$ second in the two cases of white phase noise and white frequency noise, for an operating frequency 13.8 GHz.

**Simulations**
We show here one example of white phase noise and another example of white frequency noise, which cause a coherence loss of 2% for an integration time $T = 60\ s$ and at the astronomical frequency 13.8 GHz, the highest frequency of the SKA Mid telescope. The frequency reference which has to be delivered by the frequency dissemination system is taken as 4 GHz. It is then used to sample signals in different bands. In the band of the highest frequency 13.8 GHz, the frequency reference 4 GHz will be multiplied by 8 before being applied to the A/D converter for this signal chain.

*White phase noise:*
We show in figure 5 an example of white phase noise that causes a phase deviation of RMS= 0.2 radians at 13.8 GHz and about 2% coherence loss ($L_C = 0.0198$). Figure 5 gives the phase and frequency deviation signals. Figure 6 gives the Allan deviation corresponding to this case. Figure 7 gives the coherence loss calculated directly from the phase and using a least squares fit of the Allan deviation.

*White frequency noise:*
We show in figure 8 an example of white frequency noise for a frequency reference at 4 GHz, which gives 2% coherence loss for an astronomical frequency 13.8 GHz and an integration time of T=60 seconds. Figure 8 gives the phase and frequency deviation signals. Figure 9 gives the Allan deviation corresponding to this case. Figure 10 gives the coherence loss calculated directly from the phase and using a least squares fit of the Allan deviation.

**Limitations of the calculation of the coherence loss from Allan deviation:**
We consider here two types of limitations of the calculation of the coherence loss from Allan deviation. The first limitation is due to the fact that a linear drift in the phase deviation is not reflected in the Allan deviation, while it has an effect on the coherence loss. The other difficulty is encountered when the phase deviation has a complex structure that makes a simple least squares fitting with a simple model inaccurate. This can happen, for example, in the case of a sinusoidal component in the phase deviation. A sinusoidal component in the phase deviation could be caused by vibrations or interference from other frequencies occurring within the frequency reference generator (Filler, 1988).

**Linear phase drift**
One of the limitations of using the Allan deviation in estimating the coherence loss caused by the phase noise is that the Allan deviation is insensitive to a drift in phase which, obviously, causes coherence loss. A linear phase drift is caused by an offset (DC component $y_0$) in the frequency reference, see Eq. (11). If the phase drift is of the form:

$$\varphi(t) = At \tag{31}$$

where $A = 2\pi y_0$.

Then the coherence and coherence loss are given by:

$$C(T) = \left|sinc\left(\frac{AT}{2}\right)\right|, \quad L_C = 1 - \left|sinc\left(\frac{AT}{2}\right)\right| \tag{32}$$

where $sinc(x) = \frac{\sin(x)}{x}$.

A drift $A$ of 0.1 rad/min causes a coherence loss of about 0.1 % for an integration time of about 100 s. This effect does not appear in the Allan deviation characterisation alone. This value of linear phase drift is the maximum limit acceptable for the SKA. The requirement is to limit the phase drift to less than 1 radian over 10 minutes. This requirement is related to the out-of-beam calibration.

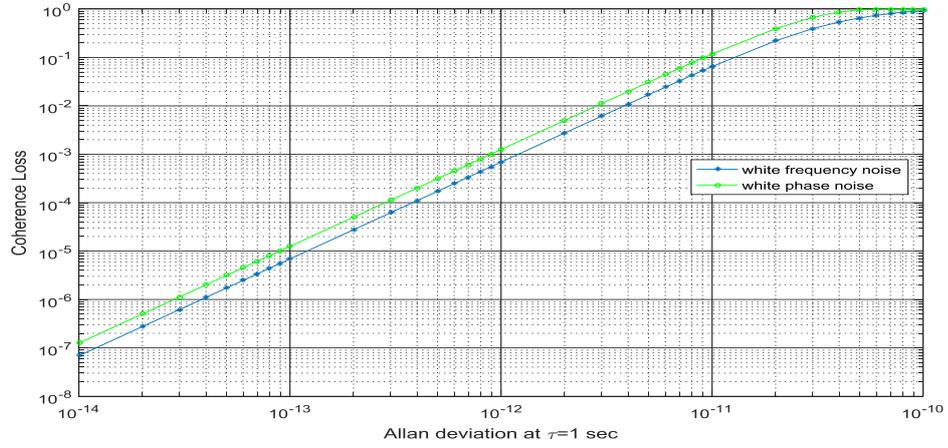

Fig. 2. The coherence loss for integration time T=1 second as function of the Allan deviation value at $\tau = 1$ second, in the two cases of white frequency noise and white phase noise, for the operating frequency 13.8 GHz.

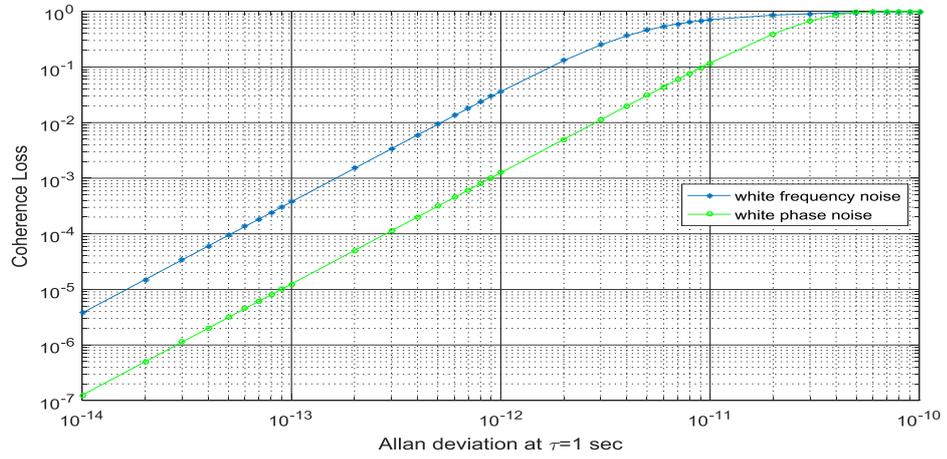

Fig. 3. The coherence loss for the integration time T=60 seconds as function of the Allan deviation value at $\tau = 1$ second, in the two cases of white frequency noise and white phase noise, for the operating frequency 13.8 GHz.

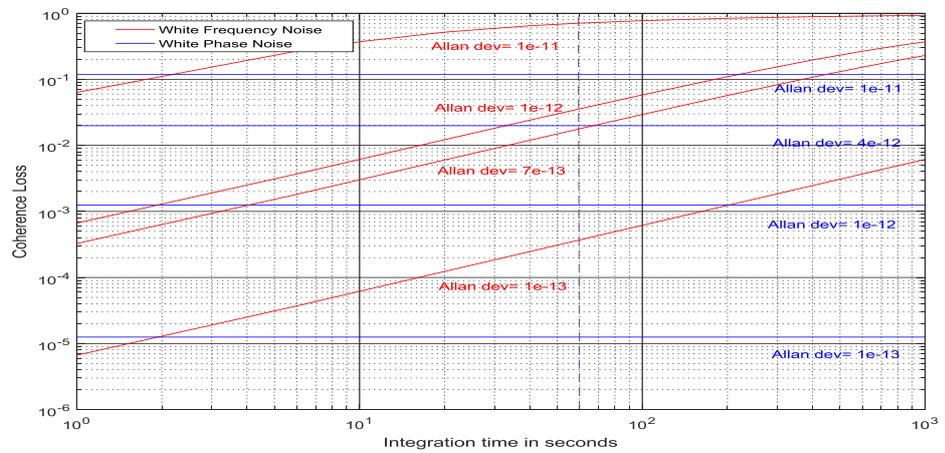

Fig. 4. The coherence loss as a function of the integration time for different values of Allan deviation at $\tau = 1\,s$, in the two cases of white frequency noise and white phase noise, for the operating frequency 13.8 GHz.

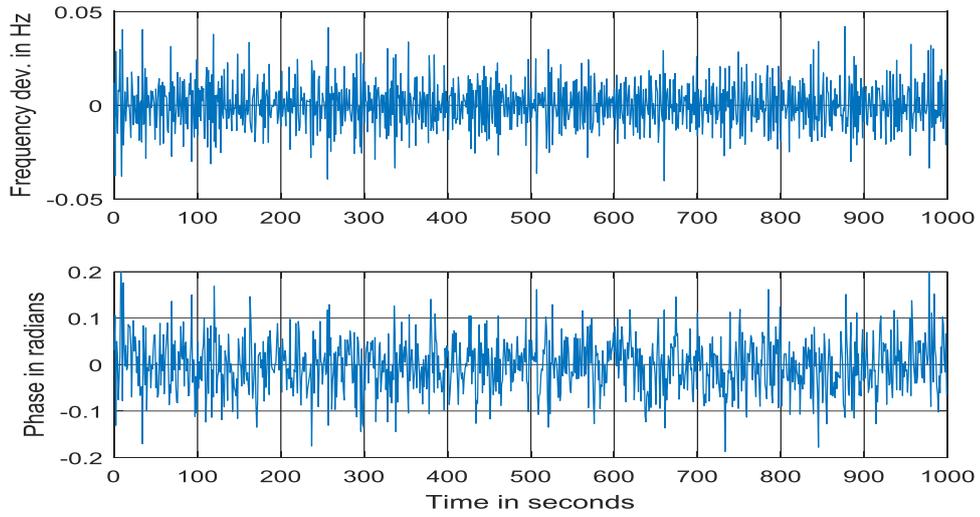

Fig. 5. Frequency deviation and phase of a white phase noise for a frequency reference at 4 GHz, which gives 2% coherence loss for an astronomical frequency 13.8 GHz.

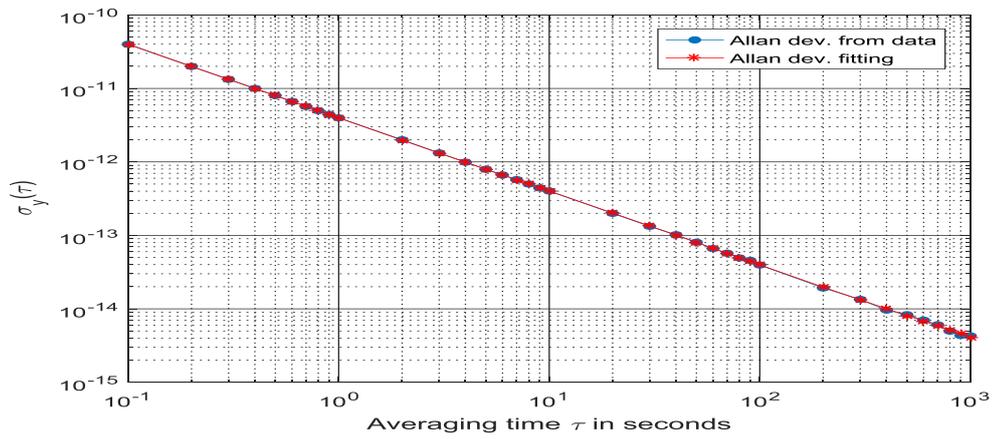

Fig. 6. Allan deviation of a white phase noise signal (the signal of figure 5) calculated from the simulation data and from a least squares fitting.

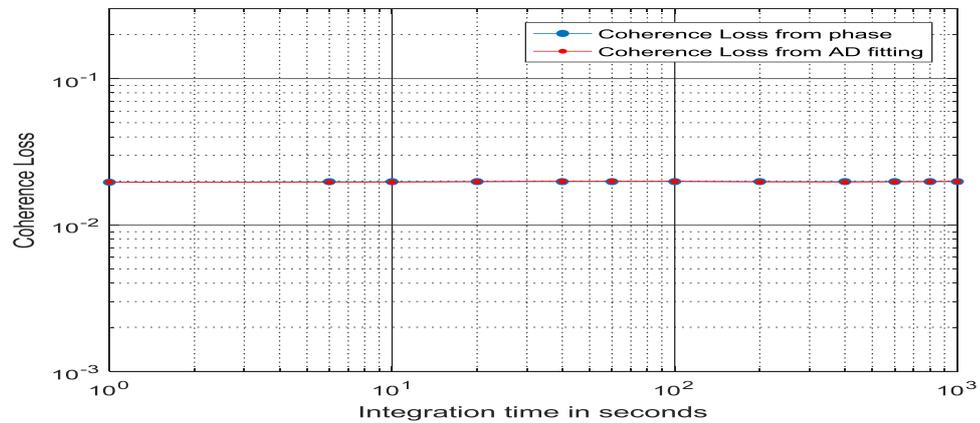

Fig. 7. Coherence Loss of the white phase noise signal of figure 5, calculated directly from the phase signal and from a least squares fitting of Allan deviation.

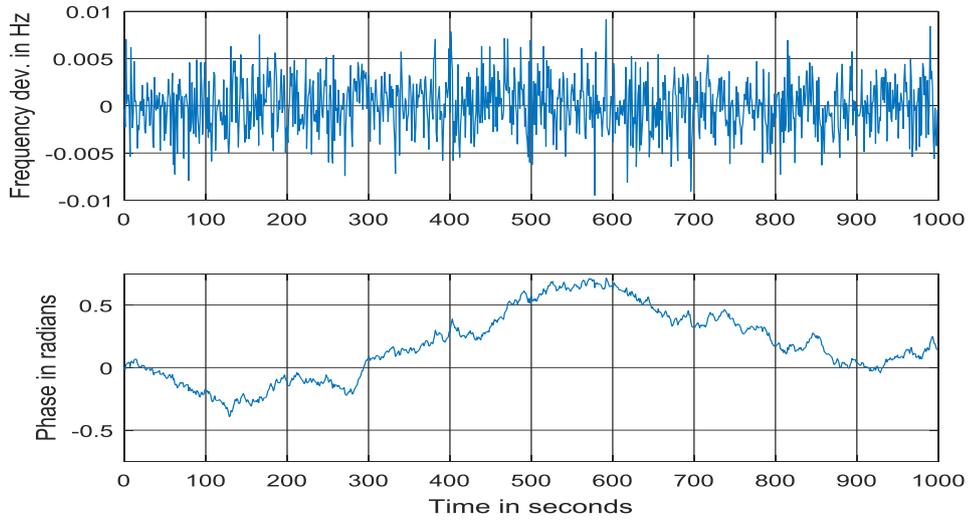

Fig. 8. Frequency deviation and phase of a white frequency noise for a frequency reference at 4 GHz, which gives 2% coherence loss for an astronomical frequency 13.8 GHz and an integration time of T=60 seconds.

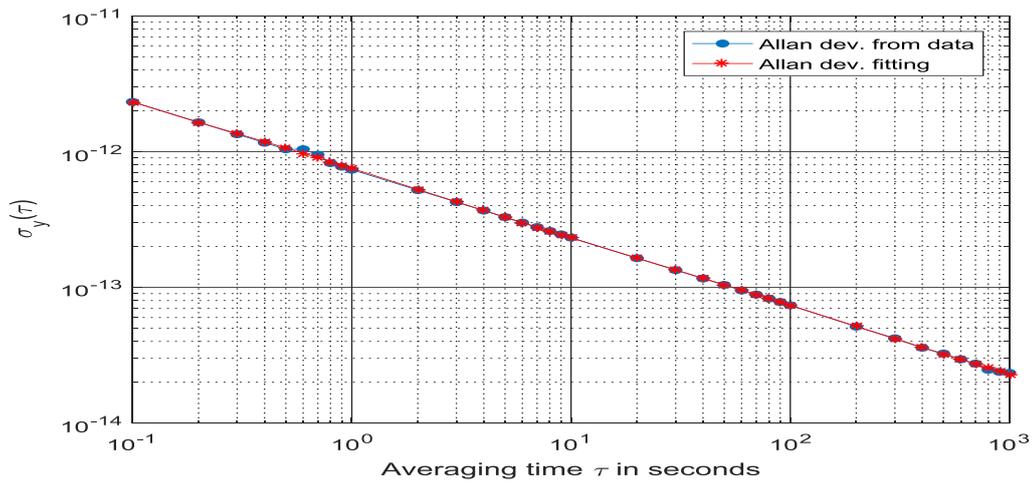

Fig. 9. Allan deviation of a white frequency noise signal (the signal of figure 8) calculated from the simulation data and from a least squares fitting.

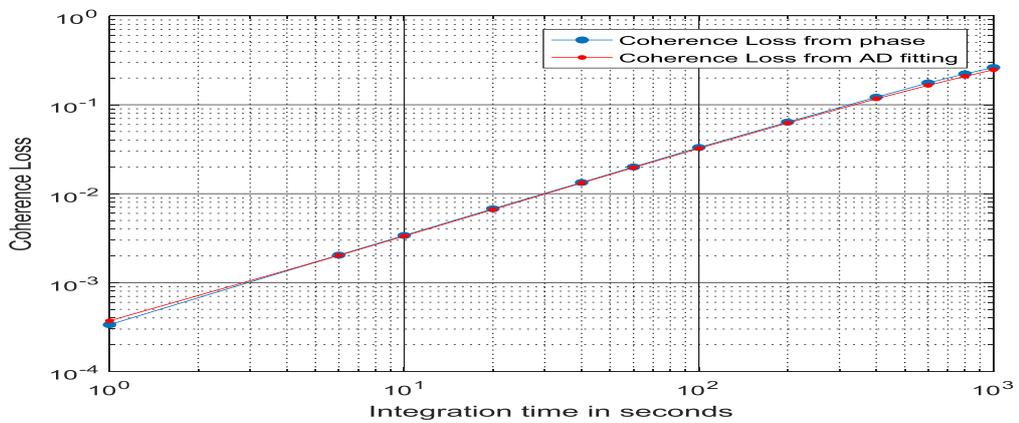

Fig. 10. Coherence Loss of the white phase noise signal of figure 8, calculated directly from the phase signal and from a least squares fitting of Allan deviation.

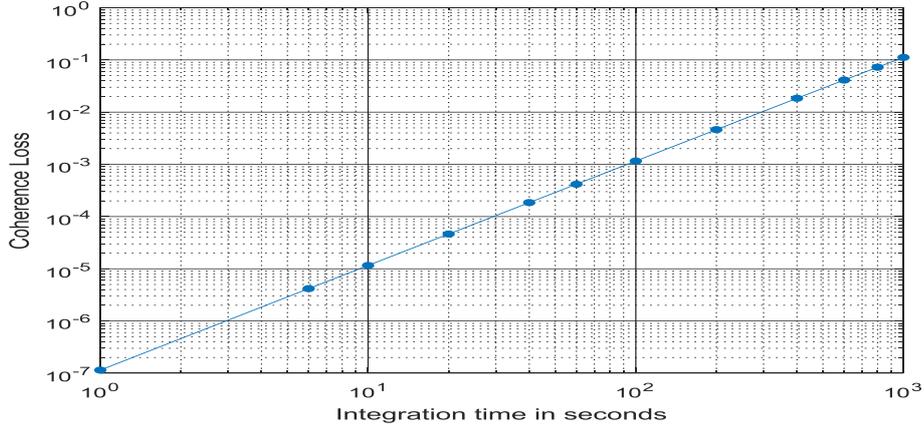

Fig. 11. Coherence loss caused by a phase drift of 0.1 rad/min.

**Sinusoidal component in the phase deviation:**

If the phase is modulated by a sinusoidal component of frequency $f_m$ then that the phase deviation includes a component of the form:

$$\varphi(t) = B\cos(2\pi f_m t) \tag{33}$$

its corresponding Allan deviation is given by:

$$\sigma_y(\tau) = B \frac{\sin^2(\pi f_m \tau)}{\pi v_0 \tau} \tag{34}$$

$v_0$ is the reference frequency.

The formula (34) can be proved using the equation (19). Replacing (33) in equation (19), we find:

$$\sigma_y^2(\tau) = \frac{1}{2} \frac{B^2}{(2\pi v_0 \tau)^2} \langle [\cos(2\pi f_m(t+2\tau)) - 2\cos(2\pi f_m(t+\tau)) + \cos(2\pi f_m(t))]^2 \rangle \tag{35}$$

$$\sigma_y^2(\tau) = \frac{2B^2}{(2\pi v_0 \tau)^2} \langle [\cos(2\pi f_m(t+\tau))\cos(2\pi f_m \tau) - \cos(2\pi f_m(t+\tau))]^2 \rangle \tag{36}$$

$$\sigma_y^2(\tau) = \frac{2B^2}{(2\pi v_0 \tau)^2} \langle \frac{(1+\cos(4\pi f_m(t+\tau)))}{2} 4\sin^4(\pi f_m \tau) \rangle \tag{37}$$

then we find equation (34) as $\langle \cos(4\pi f_m(t+\tau)) \rangle = 0$. The last term might not vanish if the instants $t$ are distributed in a way so that this term does not average to 0. The instants $t$ determine the beginning of the averaging segments in the Allan deviation calculation.

Figure 12 shows the Allan deviation curves of two phase noise signals, the first corresponding to the phase deviation of the sum of a white phase noise of 0.2 radian at 13.8 GHz and a sinusoidal component of a frequency 10 Hz and an RMS value of 0.2 radians (amplitude of $0.2\sqrt{2}$ radians) at 13.8 GH, and the second to the white noise component alone. A sinusoidal component in the phase of RMS value σ would cause the same value of coherence loss that is caused by a white phase noise of variance $\sigma^2$ (standard deviation σ). The coherence loss caused by a combination of phase noise and a sinusoidal component of σ = 0.2 radians each causes the same coherence loss as is seen with a white phase signal of $\sigma = 0.2\sqrt{2} \cong 0.2828$ radians, which is about 0.039 (see figure 1) as is shown in figure 13.

A sparse sampling of the same Allan deviation function would obscure the effect of the sinusoidal component. A calculation of the coherence loss based on a least squares fit of the Allan deviation over a sampling of a few points of the curve would result in significant error in the estimated values.

## 6. Conclusions

In this paper, we have analysed the relationship between the instability of a frequency reference used for radio frequency interferometry and the coherence loss caused by it. A particular case of the coherence requirements of the SKA was considered as an example. Theoretical and practical aspects, in addition to simulation results have been presented. We highlighted the importance of a careful analysis of the types of phase noise that could be masked in Allan deviation and will have an effect on the coherence loss.

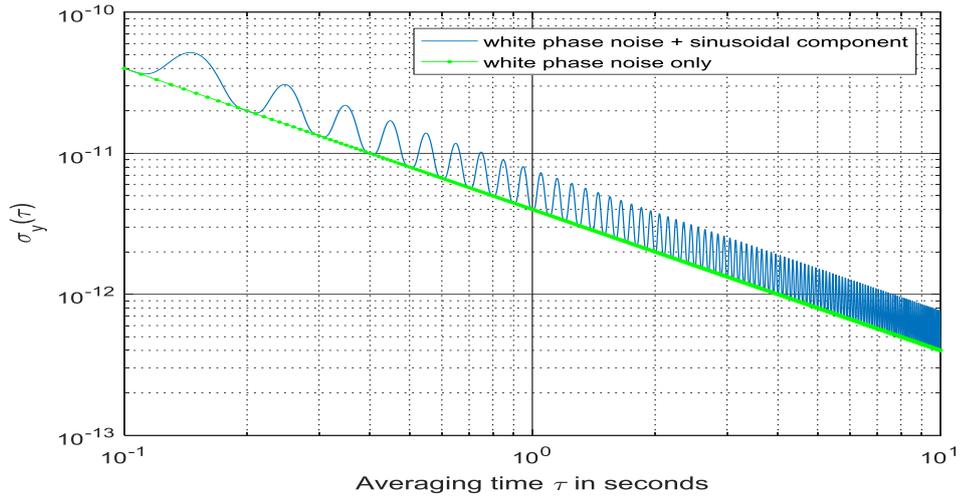

Fig. 12. Allan deviation corresponding to phase deviation of the sum of a white phase noise of 0.2 radian at 13.8 GHz and a sinusoidal component of 10 Hz and amplitude of $0.2\sqrt{2}$ radians at 13.8 GHz.

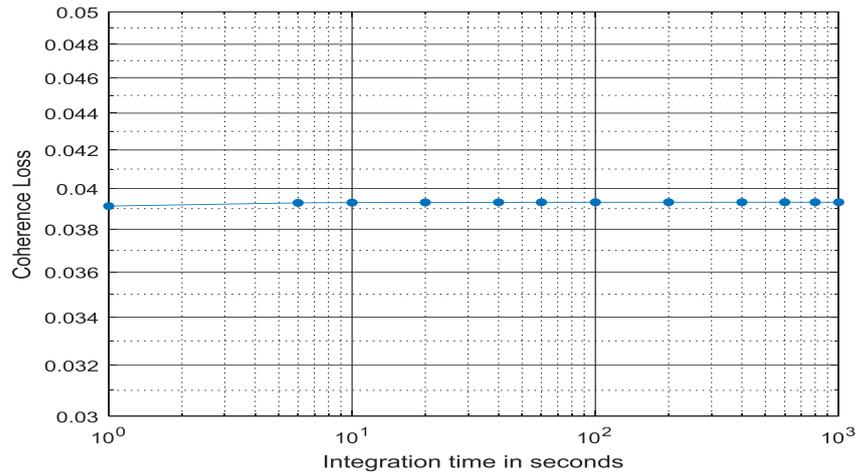

Fig. 13. The coherence loss caused by the sum of a white phase noise of 0.2 radian at 13.8 GHz and a sinusoidal component of 10 Hz and amplitude of $0.2\sqrt{2}$ radians at 13.8 GHz.

**Acknowledgements**

This work is being carried out for the SKA Signal and Data Transport (SaDT) consortium as part of the Square Kilometre Array (SKA) project. Fourteen institutions from eight countries are involved in the SaDT consortium, led by the University of Manchester. The SKA project is an international effort to build the world's largest radio telescope, led by SKA Organisation with the support of 10 member countries.